\documentclass [12pt,epsfig]{article}
\usepackage{amsmath,amssymb,epsfig}
\usepackage{graphicx}

\begin{document}
\newcommand{\el}{\left}
\newcommand{\er}{\right}
\newcommand{\dis}{\displaystyle}
\newcommand{\p}{\prime}
\newcommand{\ka}{\kappa}
\newcommand{\bc}{\beta C}
\newcommand{\rrr}{\rho}
\newcommand{\ro}{\rho^\circ}
\newcommand{\ti}{\tilde}
\begin{center}
{\Large \bf Nucleus-nucleus reaction cross-sections calculated for
realistic nuclear matter distributions within the Glauber-Sitenko approach
}\\[5mm]
{\large  V.~Lukyanov$^1$, E.~Zemlyanaya$^1$}, B.~Slowinski$^{2,3}$, \\[3mm]
{\small\it $^1$Joint Institute for Nuclear Research, Dubna, Russia~~~~~~~~}\\
{\small\it ~~~~~~~~~~~~~~~~$^2$Faculty of Physics, Warsaw University of
Technology, Warsaw, Poland}\\
{\small\it $^3$Institute of Atomic Energy, Otwock-Swierk, Poland~~~~~~~~~}\\
\end{center}
\vspace{1mm}
{\underline{\it Abstract}.}{\footnotesize Basing on the thickness (profile)
function, previously obtained for the realistic Fermi type distribution
of nucleons in nuclei, calculations are made of the microscopic eikonal
phases of the nucleus-nucleus scattering and the total reaction cross-
sections. In so doing, the phase is deduced to the one-dimensional integral
provided that the Gaussian density distribution for the projectile nucleus
and an arbitrary shape of the thickness density for the target nucleus are
used. The problems of obtaining parameters of the "point" nucleon density
are considered. A possibility of approximating the realistic densities by
the "surface-matched" Gaussian functions and the dependence of cross-sections
on the nucleon-nucleon interaction radius are discussed. The in-medium
effects and the role of the trajectory distortion are studied. Conclusions are
made on physics of the process, and comparison with experimental data is
made with cross sections calculated by using the developed method where no
free parameters are introduced.}

\vspace{1mm}
%\large
\section {Introduction}
\setcounter {equation}0

In nuclear physics, the Glauber-Sitenko approach [1,2] is used with
regard to several its modifications for investigation of nucleus-nucleus
collisions at energies of the order of 10$\div$100 MeV per nucleon of an
incident nucleus. In this case, one can determine the eikonal phase by
both introducing a model optical potential and using the microscopic
approach where it is expressed through density distributions of nuclei
and the nucleon-nucleon scattering amplitude.

The parameters of the phenomenological potential $U_{opt}(r)=V(r)+iW(r)$
are usually fitted by comparison of calculations with the experimental
differential elastic scattering and total reaction cross-sections.
Nevertheless, the problem of ambiguities of obtained parameters still
remains \cite{Sat}. For example, it was shown in \cite{LSZ} that the
total reaction cross-section obtained analytically for the Woods-Saxon
potential depends, in principle, on two combinations $RW_0$ and $R/a$
of three parameters, the radius $R$, diffuseness $a$ and the power Wo.
This fact enables one to specify these latter quantities in the wide
limits of their available values. As to the microscopic approach, it
does not include, in principle, free parameters, and enables one
to calculate the eikonal phases themselves rather than potentials of
scattering. It allows one to make predictions of the total reaction
cross-sections, including, e.g., those with participation of radioactive
nuclei, the important problem related to transmutation of radioactive
waste \cite{{W},{TW}}.

Beginning with the earlier work \cite{FST}, the microscopic approach was
applied for calculations of proton-nucleus cross-sections. Later on, based
on the respective theory of the multiple scattering of nucleons by nuclei
\cite{{G},{S}}, it was generalized in \cite{{Czyz},{Formanek}} to the
processes of nucleus-nucleus scattering. This approach was often employed
for analysis of interactions of light nuclei with nuclear targets which led,
in particular, to the discovery of the neutron hallo in $^6{He},\; ^{9}{Li}$,
the proton hallo in $^{11}{Be}$, and the establishment of nuclei with excess
of neutrons or protons (see, for instance, \cite{T}, and the review papers
\cite{{BCH},{KKF}}). Much attention was also given to studying the mechanism
of scattering of nuclei and, in particular, to effects of deflection on the
straight-line trajectory of motion \cite{{VZ},{MHC}}. The role of higher
order corrections to the eikonal phases \cite{AZW} and effects of nuclear
clusterization \cite{TJAK} were examined, too.

In the majority of such works the use of Gaussian functions (G-functions)
for nuclear densities is typical because they make it possible to separate
variables in the multidimensional integrals for eikonal phases and,
therefore, to obtain results in the analytic form. This is the main reason
why G-functions and their modifications are used in calculations of
cross-sections even for heavy ions although in this case, of course, the
physical reasons require taking the functions of extended shape. In nuclear
physics, the Fermi function (F-function) $u_F(r)=[1+exp(r-R/a)]^{-1}$ is
usually used as the most realistic function for description of densities
and potentials of medium and heavy nuclei. Unfortunately, in this case the
analytic calculations encounter difficulties. For example, it is impossible
to separate variables in the same multidimensional integrals for phases.
Nevertheless, F-functions are applied not only in numerical, but also in
analytical calculations. In the case of heavy ions these functions are
especially needed since they are most realistic ones for description of the
shape of potentials and densities in the periphery of collisions of nuclei,
the region that forms the elastic differential and total reaction
cross-sections. With reference to microscopic approaches the problem is
posed to develop analytical methods for calculating cross-sections when
using the realistic densities for nuclei with their individual parameters
being known from other experiments. In so doing, one can predict with
confidence the cross-sections for different combinations of colliding
nuclei, and thus to study the genuine mechanism of their interaction. Just
such is the goal of the present work.

In Chapter 2 presented is a series of modifications of the origin (initial)
microscopic formula for a scattering phase. This is important for both
understanding the mechanism of nucleus-nucleus scattering and calculations
free of assumptions carried out in many works. In Chapter 3, the explicit expressions
for the so-called profile functions of nuclear densities are given in the
form of the Gaussian, uniform, and symmetrized (SF) Fermi functions. It is
shown how one can reduce the 4-dimensional convolution integral to the
one-dimensional one if the density of incident (light) nuclei is presented
in the form of G-function and the density of a target nucleus in the form
of SF-function. For arbitrary forms of densities the phase is obtained
as a one-dimensional integral with the Fourier-Bessel transforms of profiles
of densities. The explicit form of such a transform is given for the profile
of the SF-density. An inference about the applicability of the so-called
modified Gaussian densities is drawn. Chapter 4, is devoted to the problems
of using nuclear densities obtained from nuclear form-factors in
electron-nuclear scattering, the effects on cross-sections of a choice
of the NN-force radius, the distortion of the trajectory of scattering, and
the in-medium effects. In Chapter 5, we give the comparison with
experimental data and general conclusions.

\section {Basic formulae}
\setcounter {equation}0

In the framework of the eikonal approximation and the microscopic approach
one can obtain the total reaction cross-sections as follows \cite{{G},{S}}:
\\
%(2.1)
\begin{equation}\label{a2}
\sigma_R=2\pi\int\limits_0^\infty db~b~\el(1- {\Huge e}^{-\dis \chi(b)}\er),
\end{equation}
where the phase
\\
%(2.2)
\begin{equation}\label{a6}
\chi(b)={\bar\sigma}_{NN}\,{\cal I}(b)
\end{equation}
is determined by the isospin average total nucleon-nucleon cross-section
\\
%(2.3)
\begin{equation}\label{d11}
{\bar\sigma}_{NN}=\frac{Z_pZ_t\,\sigma_{pp}+N_pN_t\,\sigma_{nn}+
(Z_pN_t+N_pZ_t)\,\sigma_{np}}{A_pA_t}
\end{equation}
and the folding integral that, in the case of the nucleus-nucleus scattering,
has the form \cite{{Czyz},{Formanek}}
\\
%(2.4)
\begin{equation}\label{a7}
{\cal I}(b)=\int d^2s_p\, d^2s_t~\ro_p(s_p)~\ro_t(s_t)~f(\xi),
\qquad\quad {\bf\xi}=\bf b-{\bf s}_p+{\bf s}_t.
\end{equation}
Here vectors $~{\bf s},~{\bf\xi}$ are displayed in the impact parameter
$b$ plane perpendicular to the $oz$ axis along the momentum ${\bf k}_i$
of the incident nucleus
$\footnote{~In \cite{BBS} a similar expression was obtained in the
model of interacting tubes of the nucleon fluxes in collided nuclei.}$,
and $\ro(s)$ are the so-called thickness (profile) functions of density
distributions of centers of nucleons ("point-like nucleons") of the
incident nucleus (with the atomic number $A_p$) and the target nucleus
$A_t$
$\footnote{~Here $\bf s$ and $\bf r$ are vectors in the 2- and 3-dimensional
spaces, and $r^2=s^2+z^2$. Then, $\rho(r)$ and $\ro(r)$ are the density
distributions, and  $\rho(s)$ and $\ro(s)$ are their profiles, respectively.}$.
The thickness densities are given as
\\
%(2.5)
\begin{equation}\label{a8}
\rho(s)=\int\limits_{-\infty}^\infty dz~\rho(\sqrt{s^2+z^2}).
\end{equation}
The point nucleon profiles $\ro(r)$ differ from the matter distributions
$\rho(r)$ in nuclei composed of real, i.e. "dressed" nucleons. So, when
using the convolution integral (\ref{a7}) we need be concerned with
obtaining the point-like densities $\ro(r)$ from the "experimental"
nuclear densities $\rho(r)$. Just for $\rho(r)$ a large set of tabulated data
exists obtained from electron-nucleus scattering\linebreak data
$\footnote{~In general, one-particle densities $\rho(r)$ depend on coordinates
in the respective center-of-mass frame of a nucleus. However, as usual, in
analysis of form factors of nuclei, one omits the respective factor of the
center-of-mass motion $\exp({q^2\langle q^2 \rangle}/6A)$, where ${\langle
r^2\rangle}$ and $A$ are the root-mean-square radius and a mass number of
the nucleus. Therefore, the tabulated $\rho(r)$ appear to be distributions
of a nuclear charge (or matter) in the field of a fixed nuclear potential.
At small $q$ and large $A$ densities in both systems coincide.}$,
and our goal is to develop the approach in such a way that to use in
calculations the table data for $\rho(r)$ and do not introduce
free parameters.

The function $f(\xi)$ determines the form of the nucleon-nucleon
interaction amplitude
\\
%(2.6)
\begin{equation}\label{a9}
f(\xi)=(\sqrt{\pi}a_N)^{-2}~{\Huge e}^{\dis-\xi^2/a_N^2},
\qquad\qquad\quad     a_N^2={2\over 3}\,r_{N\,rms}^2.
\end{equation}
Here $r^2_{N\,rms}$ is the root-mean-square radius of NN-interaction,
and $a^2_N=2\beta$ is expressed by the shape parameter $\beta$ of the
scattering amplitude of nucleons$\footnote{~The amplitude is $f_N(q)=
f_{N}(0)\, f(q)$, where $f_{N}(0)=(k_{N}/4\pi){\bar\sigma}_{NN}(i+
\alpha_{NN})$, and $k_N$ is the relative momentum of colliding nucleons.
For $f(q)=\exp(-q^2a_N^2/4)$ the Bessel-Fourier transform $f(\xi)=
(2\pi)^{-2}\int \exp(-i{\bf q}{\bf\xi}) \,f(q)d^2q$ follows eq.(\ref{a9}),
and in the zero-range approximation ($a_N=0$), when $f(q)=1$, one gets the
delta-function in 2-dimensional space, i.e. $f(\xi)=\delta^{(2)}(\xi)$.}$
in the form $\exp(-q^2\beta/2)$. The values of $\beta$  at energy of
the order of 1 GeV are in the interval 0.21$\div$0.23 fm$^{-2}$ \cite{AAV}
which means that $r^2_{N\,rms}=0.63 \div 0.69$ fm$^2$. In our case the
nucleon-nucleon forces act in nuclear medium, and to take into account their
influence the correction factor $f_m$ is additionally introduced under the
integral. Later on we will touch this problem in detail.

The convolution integral (\ref{a7}) has a similar form as the 6-dimensional
"double folding" integral in calculation of the nucleus-nucleus potential
\cite{SL}. In both the cases we are led to search for the ways of separating
variables in integrands. In Chapter 3 we show that the integral (\ref{a7})
can be calculated explicitly if both densities are the G-functions, or it is
reduced to the one-dimensional integral when one of the densities has the
Gaussian form.  At the same time, there exists a standard procedure to
transform such integrals to one-dimensional ones in the momentum space.
To this end, in each function in (\ref{a7}) one should perform the
two-dimensional Fourier-Bessel transformation
\\
%(2.7)
\begin{equation}\label{a10}
u(s)={1\over (2\pi)^2}\int \, e^{-\dis{i{\bf k}{\bf s}}} \,{\ti u}(k)d^2k =
{1\over 2\pi}\int\limits_0^\infty \, J_0(ks) \, {\ti u}(k)kdk,
\end{equation}
where
%(2.8)
\begin{equation}\label{a11}
{\ti u}(k)= \int \, e^{\dis{i{\bf k}{\bf s}}} \,u(s)d^2s =
2\pi \int\limits_0^\infty \, J_0(ks)\, u(s)sds.
\end{equation}
Then (\ref{a7}) becomes
%(2.9)
\begin{equation}\label{a12}
{\cal I}(b)={1\over 2\pi}\int\limits_0^\infty kdk~J_0(kb)~
{\ti \ro}_p(k)~{\ti \ro}_t(k)~{\ti f}(k),
\end{equation}
where
%(2.10)
\begin{equation}\label{a13}
{\ti f}(k)=\exp\el(-k^2r_{N\,rms}^2/ 6\er).
\end{equation}
Next, using the convolution formula for the nuclear thickness density
%(2.11)
\begin{equation}\label{a14}
\rho_i(s)=\int d^2s_N\,\rho_N(s_N)\,\ro_i(|{\bf s}-{\bf s}_N|),
\end{equation}
where $\rho_N(s_N)$ is the nucleon thickness density, one obtains with
the help of (\ref{a10}) the following result:
%(2.12)
\begin{equation}\label{a15}
\ti{\rho}_i(k)={\ti\rho}_N(k)\,{\ti\ro}_i(k).
\end{equation}

For the Gaussian density of a nucleon with the $rms$ radius one has
%(2.13)
\begin{equation}\label{a16}
\ti{\rho}_N(k)\,=\,\exp\Bigl(-{k^2r^2_{0\,rms}\over 6}\Bigr).
\end{equation}
Then, (\ref{a12}) results in
%(2.14)
\begin{equation}\label{a17}
{\cal I}(b)={1\over 2\pi}\int\limits_0^\infty kdk \, J_0(kb)\,
{\ti \ro}_p(k) \, {\ti \rho}_t(k) \, \exp\Bigl(-{k^2\tau^2\over 6}\Bigr),
\end{equation}
%(2.15)
\begin{equation}\label{a18}
\tau^2\,=\,r^2_{N\,rms}-r^2_{0\,rms}.
\end{equation}
If it is considered that  $r^2_{N\,rms}$ and $r^2_{0\,rms}$ coincide,
then $\tau^2=0$ and so
%(2.16)
\begin{equation}\label{a19}
{\cal I}(b)={1\over 2\pi}\int\limits_0^\infty kdk \, J_0(kb)\,
{\ti \ro}_p(k) \, {\ti \rho}_t(k),
\end{equation}
and in the coordinate representation
%(2.17)
\begin{equation}\label{a20}
\qquad {\cal I}(b)=\int\limits_0^\infty d^2s \,
\ro_p(|{\bf b}-{\bf s}|) \, \rho_t(s).
\end{equation}
As a result, we have obtained the expressions for convolution integrals
(\ref{a17}),(\ref{a19}),(\ref{a20}), where the thickness functions
$\rho_t (s)$ (or $\ti\rho_t (k)$) come into being instead of profiles of the
target-nucleus "point-like" densities. Just for them the respective
"experimental" densities are usually known from tables where they are
parametrized for heavy and middle-weight nuclei in the form of a Fermi
function. In principle, one can employ eqs.(\ref{a15}) and (\ref{a16}) for
incident nuclei as well, i.e. one uses $\ti\ro_p=\ti\rho_p/\ti\rho_N$. But
then under the integrals (\ref{a17}),(\ref{a19}) an increasing Gaussian
function appears, and if for the profiles of both the densities the functions
with realistic exponential asymptotics are taken, then the integrals will
diverge at the upper limit. Of course, one can act formally, i.e. either
"cut" integration at the point where an integrand starts to increase,
or change the Gauss-like nucleon form-factor (\ref{a16}) by a dipole
formula (see below eq.(\ref{d2})). On the other hand, if for the density
of one of the nuclei one takes a Gauss function, then no divergence arises.

Finally, it should be mentioned that in some papers the so-called
zero-range approximation ($r^2_{N\,rms}=0$) is used. This leads to the
convolution integral in the form (\ref{a12}) with $\tilde f(k)$=1, or,
in the spatial coordinates, in the form (\ref{a20}) with $\ro_t(s)$
in(\ref{a20})stead of $\rho_t(s)$. A rougher approach is when both the
densities in (\ref{a20}) are supposed to be the nuclear ones
$\rho(s)$. Now we see that such approaches are not necessary to be used, and
they themselves distort the true mechanism of nucleus-nucleus scattering.

\section {Eikonal phases for realistic density distributions}
\setcounter {equation}0

  As mentioned above, to obtain analytic expressions for phases and
cross-sections the Gaussian (G-functions) density distributions and their
profiles are used
%(3.1)
\begin{equation}\label{c1}
\rho_G(r)=\rho_G(0)~{\Huge e}^{\dis-{r^2/a_G^2}}, \qquad \qquad \qquad
\rho_G(0)=A/({\sqrt\pi}a_G)^3,
\end{equation}
%(3.2)
\begin{equation}\label{c2}
\rho_G(s)=({\sqrt\pi}a_G)~\rho_G(0)~{\Huge e}^{\dis-{s^2/a_G^2}},
\qquad\qquad\quad     a_G^2={2\over 3} R_{rms}^2,
\end{equation}
where the only parameter $a_G$ is determined by the root-mean-square nuclear
radius $R_{rms}$
$\footnote{~For the point density $\ro_G(r)$, the parameter
$a^{\circ\,2}_G={2\over 3} {\cal R}_{rms}^2$ can be expressed through
the respective $rms$-radius ${\cal R}^2_{rms}=R_{rms}^2 \, - \, r_{0\,rms}^2$.}$.

We also list the functions of uniform density distribution and the
relevant profiles
{\samepage
%(3.3)
\begin{equation}\label{c3}
\rho_u(r)=\rho_u(0)~\Theta(R_u-r), \qquad\qquad\qquad
\rho_u(0)=3A/4\pi R_u^3,
\end{equation}
%(3.4)
\begin{equation}\label{c4}
\rho_u(s)=\rho_u(0)\sqrt{R_u^2-s^2}~\Theta(R_u-s), \qquad\qquad
R_u^2={5\over 3} R_{rms}^2,
\end{equation}
}
which are sometimes used for middle and heavy nuclei.

In principle, the realistic density having the Fermi distribution
(F-function) can be approximated by a sum of Gaussian functions
with by fitting expansion coefficients and parameters $a_G$. Such
a procedure was suggested in \cite{DK}, and in \cite{Rihan} this
one was proposed not to densities but directly to profiles of
Fermi functions. Unfortunately, this procedure must be repeated at
every new set of parameters $R$ and $a$ of the initial
F-functions. However, one can remind that for heavy ions, both
differential elastic scattering and total reaction cross-sections
are first determined by the behavior of phases in a periphery of
collisions. It seems likely that the first attempt to model the
tail of the F-distribution $\rho_F(r)$ with the help of $\it one$
Gaussian function was made in \cite{Ka}. Later on in \cite{CG},
not a Fermi density but its profile $\rho_F(s)$ was reproduced in
such a way. For this aim the G- and F-functions were matched in
their periphery to find two parameters $a_{\bar G}$ ¨ $\ro_{\bar
G}(0)$ of the so-called modified G-function
\\
%(3.5)
\begin{equation}\label{c5}
\rho_{\bar G}(s)=({\sqrt\pi}a_{\bar G}~)~\rho_{\bar G}(0)~{\Huge e}^
{\dis-{s^2/a_{\bar G}^2}}.
\end{equation}
This function is not normalized, since its parameters are no longer connected
mutually in such a way as in (\ref{c1}). In the general case, when matching $\rho_{\bar G}(s)$
with an arbitrary form of the extended function $\rho(s)$ at two points
$s_1$ and $s_2$ we obtain two parameters of (\ref{c5})
\\
%(3.6)
\begin{equation}\label{c6}
\rho_{\bar G}(0)=\el({\sqrt\pi}\, a_{\bar G}\er)^{-1}\rho(s_1)~
\exp(s_1^2/a_{\bar G}^2),
\end{equation}
%(3.7)
\begin{equation}\label{c7}
a_{\bar G}=\Bigl[{s_2^2-s_1^2\over \ln\rho(s_1)-\ln\rho(s_2)}\Bigr]^{1/2}.
\end{equation}
In \cite{CG}, the points of matching the $\bar G$- and F-profiles have
been taken as $s_1=c$ and $s_2=c+4d$ where $c$ and $d$ are the radius and
diffuseness parameters, and the profiles $\ro_{F}(s)$ were obtained by
numerical integration in (\ref{a8}). Herewith $d$ was assumed to be the
same $d=0.53~fm$ for all nuclei, and $c$ was determined from the known
$rms$ radii of nuclei and nucleon $R_{rms}$ and $r_{0\,rms}$ with the
help of the formula
\\
%(3.8)
\begin{equation}\label{c8}
{\cal R}^2_{rms}=R_{rms}^2  - r_{0\,rms}^2 ={3\over 5}c^2\Bigl[1+{7\over 3}
\Bigl({\pi d\over c}\Bigr)^2\Bigr].
\end{equation}
In general, the accuracy of such a matching should be checked every
time since the nuclear Fermi densities $\rho_{F}(r)$ have different
values of $a$ for different nuclei. Besides, the obtained parameters
of $\bar G$-functions depend on a choice of matching points.

Starting with \cite{ELP} the symmetrized SF-function
\\
%(3.9)
\begin{equation}\label{c9}
u_{SF}(r)=\frac{\sinh R/a}{\cosh R/a+\cosh r/a}=
\frac{1}{1+\exp{r-R\over a}} - \frac{1}{1+\exp{r+R\over a}}. \qquad
\end{equation}
has come into use, first, in calculations of nuclear form factors in
eA-scattering, and next, in other problems of nuclear physics, too.
This function has several advantages as compared to the F-function,
allowing much room for analytical calculations \cite{SM}, \cite{GKLS}.
Its shape is the universal one for a satisfactory modelling the nuclear
densities of light, middle and heavy nuclei \cite{BKLP}. It is evident
from (\ref{c9}) that for middle and heavy nuclei ($R\gg a$) this function
in fact coincides with the usual Fermi function $u_F(r)=1/(1+\exp[(r-R)/a])$.
Therefore, parameters of this function can be taken from the existing
Tables of Fermi distributions for both nuclear densities \cite{VJV} and
point-like densities of nuclei \cite{BL}. For our task it is important
that just for the SF-function the respective profile was obtained in an
explicit form \cite{LZ} and therefore the following calculations can be
considerably simplified. So, the SF-density distribution and its
profile have the following form (3.10) (3.11):
\\
%(3.10)
\begin{equation}\label{c10}
\rho_{SF}(r)=\rho_{SF}(0)~\frac{\sinh R/a}{\cosh R/a+\cosh r/a}, \qquad\quad
\rho_{SF}(0)=\frac{3A}{4\pi R^3}\Bigl[1+\el(\frac{\pi a}{R}\er)^2\Bigr]^{-1},
\end{equation}

%(3.11)
\begin{equation}\label{c11}
\rho_{SF}(s)=2R~\rho_{SF}(0)~\frac{\sinh R/a}{\cosh R/a+\cosh s/a}~P(s).
\qquad\qquad\qquad
\end{equation}
\noindent
Here the main dependence of the profile on $s$ is determined by SF-function
with the same parameters as in the density $\rho_{SF}(r)$. The corrective
factor $P(s)$ is presented in \cite{LZ} and specified with the help of
the auxiliary function $x(s)$. This latter obeys the condition $x(s)\ll 1$
which allows one to simplify $P(s)$, so that it arrives at
\\
%(3.12)
\begin{equation}\label{c12}
P(s)={a\over R}~\ln(4/x(s)), \qquad
x(s)={2\over \kappa}~\frac{\cosh s/a}{\cosh s/a+\cosh R/a}
\el\{1+{\kappa -1\over\cosh s/a}\er\}.
\end{equation}
\\
Here $\kappa$ is expressed by the radius $R$ and diffuseness $a$ as
\\
%(3.13)
\begin{equation}\label{c14}
\kappa={\Huge e}^{\dis\delta}, \qquad \delta=1.10315+0.34597(R/a)-0.00446(R/a)^2.
\end{equation}
The numerical values of the coefficients in (\ref{c14}) were
found in \cite{LZ} by fitting the profile (\ref{c11}) to numerical values
of the profile integral (\ref{a8}) for $\rho_{SF}(r)$ (\ref{c10}) in the
region of $5\leq R/a \leq 20$. In the center of a nucleus one has $P(0)=1$,
and in the region of the main contribution from $s=R$ to $\infty$ it
changes a little by $\simeq 0.4(a/R)$. This enables one to take $P(s)$
at one point only, for example, at $s=R$, or at $s=s_{1/3}= R+a\ln 2$,
where the density itself falls by three times
$\footnote{~It was established in \cite{SL} that the behavior of the
nucleus-nucleus scattering is responsible for the region where densities
overlap in their periphery at $s\geq s_{1/3}$.}$. Then, if
$\cosh~R/a\gg\kappa$, one obtains
\\
%(3.14)
\begin{equation}\label{c15}
P(R)\simeq {a\over R}\el[\ln{ 4\kappa}\er]={a\over R}\el[2.48945+
0.34597{R\over a}-0.00446 \el({R\over a}\er)^2\er],
\end{equation}

and, respectively,
\\
%(3.15)
\begin{equation}\label{c16}
\rho_{SF}(s)\simeq 2R~\rho_{SF}(0)~\frac{\sinh R/a}{\cosh R/a+\cosh s/a}~
P_a(R).
\end{equation}
\\

%%%%%%%%%%%%%%%%%%%%%%%%%%%%%%%%%%%%%%%%%%%%%%%%%%%
\begin{figure}
\begin{center}
\epsfig{file=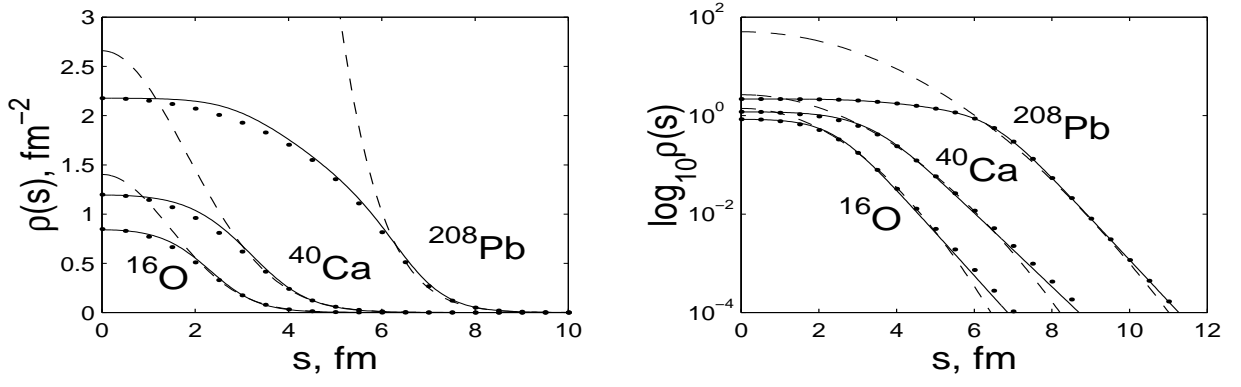,height=5cm,width=.95\linewidth}
\end{center}
\caption{Thickness functions for SF-densities. Bold dots --
numerical integration of eq.(2.5), dashes -- $\bar G$-profiles
adjusted to the bold dots curves. Solid curves -- SF-thickness
densities calculated by analytic eq.(3.11). Parameters are in
Table 1.}
\end{figure}
%%%%%%%%%%%%%%%%%%%%%%%%%%%%%%%%%%%%%%%%%%%%%%%

In Fig.1 are shown the profiles of SF-density for three nuclei:
$^{16}O,~^{40}Ca,~^{208}Pb$, calculated numerically (points) using
eq.(\ref{a8}), and the corresponding $\bar G$-profiles (dashes),
matched according to (\ref{c6}), (\ref{c7}) at $R$ and $R+4a$.
Full curves represent $\rho_{SF}(s)$ calculated with the help of the
analytic formula (\ref{c11}). The parameters of SF-densities are
taken from \cite{BKLP} and are given in Table 1. One can see that
a strong discrepancy exists between $\bar G$-profiles and the
initial SF-profiles in the inner region (for example, by two
orders of magnitude for $^{208}Pb$), and also in the region where
the density falls off by two orders of magnitude and more. The
profiles of the uniform and Gaussian forms differ much more from
the profiles of the Fermi function. In Fig.2, the functions
$\rho_{SF}(s)$ are depicted for the same nuclei but calculated
within analytic formulae: full - by (\ref{c11}) with the exact
corrective factor $P(s)$, and dashed lines - by eq. (\ref{c16})
with the approximated one $P_a(s=R)$. It turns out that the use of
the corrective factor at the radius point practically does not
change the behavior of profile functions in the peripheral region.
A slight difference, no more than a factor of two for $^{208}Pb$,
appears in the inner region only, which is far less than it is
observed in Fig.1 when one uses the $\bar G(s)$-functions. \\

{ { {\bf Table 1.}~~{\it Parameters of nuclear symmetrized Fermi
density distributions $\rho_{SF}(r,R,a)
$$\footnote{~Parameters of the Fermi-distributions
$\rho_{F}(r,R,a)$, taken from  \cite{VJV}, are very close to those
of $\rho_{SF}(r,R,a)$ for the given three nuclei}$.}  } {\large
\begin{center}
\begin{tabular}{|l|c|c|c|c|}
\hline \hline
~ Nucleus~   &~$R,~fm$~& ~$a,~fm$~ &~$R_{rms},~fm$~ & ~Ref.~\\
\hline \hline
~  $^{12}$C   & 2.214 & 0.488 &  2.496 & ~\cite{BKLP}~\\

~  $^{16}$O   & 2.562 & 0.497 &  2.711 & ~\cite{BKLP}~\\

~  $^{20}$Ne  & 2.74  & 0.572 &  3.004 & ~\cite{VJV}~\\

~  $^{24}$Mg  & 2.934 & 0.569 &  3.105 & ~\cite{BKLP}~\\

~  $^{27}$Al  & 3.07  & 0.519 &  3.06 & ~\cite{VJV}~\\

~  $^{28}$Si  & 3.085 & 0.563 &  3.175 & ~\cite{BKLP}~\\

~  $^{32}$S   & 3.255 & 0.601 &  3.370 & ~\cite{BKLP}~\\

~  $^{40}$Ca  & 3.556 & 0.578 &  3.493 & ~\cite{BKLP}~\\

~  $^{66}$Zn  & 4.340 & 0.559 &  3.952 & ~\cite{VJV}~\\

~  $^{89}$Y   & 4.86  & 0.542 &  4.27  & ~\cite{VJV}~\\

~  $^{208}$Pb & 6.557 & 0.515 &  5.427 & ~\cite{BKLP}~\\
\hline
\end{tabular}
\end{center}
}

%%%%%%%%%%%%%%%%%%%%%%%%%%%%%%%%%%%%%%%%%%%%%%%%%%%
\begin{figure}
\begin{center}
\psfig{file=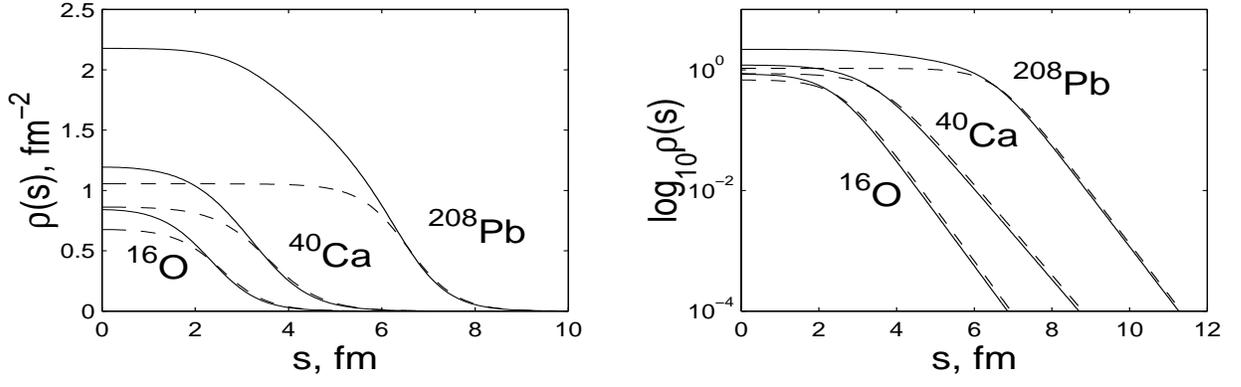,height=5cm,width=.95\linewidth}
\end{center}
\caption{Comparison of the SF-thickness densities calculated by
analytic eq.(3.11) with the exact correction function P(s) (solid
curves), with the approximate P(s=R) by eq.(3.15). Parameters are
the same as in Fig.1.}
\end{figure}
%%%%%%%%%%%%%%%%%%%%%%%%%%%%%%%%%%%%%%%%%%%%%%%

So, in the subsequent discussion we have a possibility of choosing the
profiles $\rho(s)$ of the explicit analytic form for different nuclear
densities, namely, for the Gaussian, uniform and symmetrized Fermi
functions. Below they will be used in calculations of microscopic eikonal
phases $\chi(b)$ and total reaction cross-sections.

First we consider the convolution integral (\ref{a20}) where the
thickness density of an incident nucleus $\ro_p(\zeta)$ is taken in
the Gaussian form (\ref{c5}). If one substitutes into (\ref{a20}) the
expression
%(3.16)
\begin{equation}\label{á17}
\ro_{\bar G,p}(|\vec b -\vec s|)=({\sqrt\pi}a^\circ_{\bar G,p})~
\ro_{\bar G,p}(0)~\exp\Bigl[-{1\over a^{\circ~2}_{\bar G,p}}
\el(b^2~-~2bs\cos\varphi~+~s^2\er)\Bigr]
\end{equation}
and integrates over $\varphi$ by using the definition of the Bessel function
$I_0(x)$ \cite{RG} of an imaginary argument, it follows:
%(3.17)
\begin{equation}\label{c18}
{\cal I}_{\bar G,t}(b)=2\pi(\sqrt{\pi} a^\circ_{\bar G,p})\ro_{\bar G,p}(0)
\exp\Bigl(-{b^2\over a^{\circ~2}_{\bar G,p}}\Bigr)
\int\limits_0^\infty s ds ~\rho_t(s)~
\exp\Bigl(-{s^2\over a^{\circ~2}_{\bar G,p}}\Bigr)
~I_0\Bigl({2bs\over a^{\circ~2}_{\bar G,p}}\Bigr).
\end{equation}
In a more general case, when the magnitudes of $rms$ radii of the nucleon
and NN-interaction differ, it is convenient to use the convolution integral
(\ref{a7}) with profiles of point-like densities for both nuclei. Then
one can show that in the case of the $\bar G$-thickness density of an
incident nucleus (\ref{c5}) the convolution integral takes the form:
%{\samepage
%(3.18)
$$
{\cal I}_{\bar G,N,t}(b)=2\pi\frac{a^{\circ\,2}_{\bar G,p}}
{a^{\circ\,2}_{\bar G,p}+a_N^2}(\sqrt{\pi}a^\circ_{\bar G,p})
\,\ro_{\bar G,p}(0)\,\exp\Bigl(-{b^2\over a^{\circ~2}_{\bar G,p}+a_N^2}\Bigr)
\times
\qquad\qquad\qquad\qquad \qquad\qquad\qquad
$$
\begin{equation}\label{c19}
\qquad\qquad \qquad\qquad
\times\int\limits_0^\infty sds~\ro_t(s)~\exp\Bigl(-{s^2\over
a^{\circ~2}_{\bar G,p}+a_N^2}\Bigr)~
I_0\Bigl({2bs\over a^{\circ~2}_{\bar G,p}+a_N^2}\Bigr).
\end{equation}
\\
%}
If for a target-nucleus one also takes the Gaussian function, then
integration in (3.19) is performed explicitly \cite{RG}, and one obtains
\cite{Ka}
%(3.19)
\begin{equation}\label{c20}
{\cal I}_{\bar G,N,\bar G}(b)={1\over \pi}\frac{(\sqrt{\pi}a^\circ_{\bar G,p})^3
(\sqrt{\pi}a^\circ_{\bar G,t})^3}{a^{\circ~2}_{\bar G,p}+
a^{\circ~2}_{\bar G,t}+a_N^2}\, \ro_{\bar G,p}(0)\ro_{\bar G,t}(0)\, \exp
\Bigl(-{b^2\over a^{\circ~2}_{\bar G,p}+a^{\circ~2}_{\bar G,t}+a_N^2}\Bigr).
\end{equation}
\noindent
Note that in the case of normalized Gaussian functions (\ref{c1}) it is
necessary to change ${\bar G}$ by $G$ in (\ref{c18})\,-\,(\ref{c20}), and
put $({\sqrt{\pi}}a^\circ_i)^3 \,\ro_{G,i}(0)=A_i$.

We imply that the realistic density distributions of middle and heavy nuclei
are the (symmetrized) Fermi functions. Their profile functions (\ref{c10})
and (\ref{c16}) are known in a certain form. The Bessel functions $I_0(x)$
and $J_0(x)$ are also known explicitly in the form approximated by
polynomials \cite{AS}. So all functions in the convolution integrals
(\ref{c18}) and (\ref{c19}) are given explicitly which is highly feasible
for numerical integration.

In the case when the density distributions of both the nuclei are given as
the SF-functions, it is reasonable to use the convolution integrals in
the momentum representation (\ref{a12}), (\ref{a17}) or (\ref{a19}).
Then, it is convenient to take their thickness functions approximated by
(\ref{c16}); as a result, the Fourier transform can be easily calculated.
Indeed, inserting (\ref{c16}) into (\ref{a11}) one obtains
%(3.20)
\begin{equation}\label{c21}
\ti\rho_{SF}(k)\,=\,4\pi R~\rho_{SF}(0)~P_a(R)~{\cal F}_{SF}(k, a, R),
\end{equation}
%\noindent
where
%(3.21)
\begin{equation}\label{c22}
{\cal F}_{SF}(k, a, R)\,\equiv\,{\cal F}_{SF}(k)\,=\,\int\limits_0^\infty \,
sdsJ_0(ks)~\frac{\sinh R/a}{\cosh R/a+\cosh s/a}.
\end{equation}
Taking into account the peripheral nature of nucleus-nucleus collisions
one can assume that the main contribution is made in the region when $ks\gg 1$.
Then (see, e.g., \cite{GKLS2}) we have
%(3.22)
\begin{equation}\label{c23}
{\cal F}_{SF}(k)\,=\,{\pi aR\over\sinh{\pi ak}}\,J_1(kR).
\end{equation}
In \cite{SM}, the higher order corrections to (\ref{c23}) are established
but they do not give significant contributions at $kR\gg 1$.

%%%%%%%%%%%%%%%%%%%%%%%%%%%%%%%%%%%%%%%%%%%%%%%%%%%
\begin{figure}
\begin{center}
\psfig{file=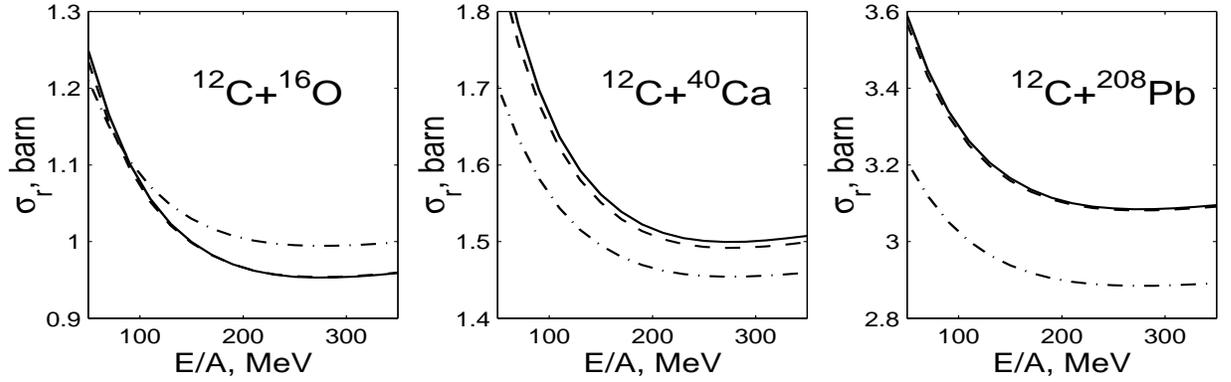,height=5cm,width=.95\linewidth}
\end{center}
\caption{Total reaction cross-sections for different thickness
densities. The $\bar G$-profile of the projectile nucleus $^{12}C$
is adjusted to its SF-point-like thickness density having
parameters from Table 2. For the target-nuclei, solid curves are
for the SF-densities (parameters in Table 1), dashes-dots -- for
unified densities, dashes -- $\bar G$-model.}
\end{figure}
%%%%%%%%%%%%%%%%%%%%%%%%%%%%%%%%%%%%%%%%%%%%%%%

    In Fig.3 are shown the calculated total reaction cross-sections
for collisions of $^{12}C$ with $^{16}O,~^{40}Ca,~^{208}Pb$ at
energies from 50 to 350 MeV/nucleon. For the target-nuclei,
different nuclear densities are chosen in the form of
SF-functions, modified ${\bar G}$-functions and U-functions of
uniform distribution. The convolution integral (\ref{c18}) has
been used for the $\chi$-phase. The profile function $\bar G$ for
the incident nucleus $^{12}C$ was shifted at $s_1=c$ ¨ $s_2=c+4a$
(see (3.5)-(3.7)) with the profile of the corresponding point-like
SF-density whose parameters are given in Table 2. When one tested
$\bar G$-profiles for the target-nuclei, the parameters of their
densities $\rho_{SF}$ were taken from Table 1. The $R_{rms}$-radii
which have been used in calculations of the uniform distribution
radius $R_u$ (\ref{c4}) are also quoted in Table 1. The
energy-depended total nucleon-nucleon cross-sections $\sigma_{NN}$
are taken from \cite{CG}. One can see that for a uniform density,
the total reaction cross-sections (dot-and-dash curves) have
highly different forms as compared to ones for the physically
justified SF-densities (full lines). Besides, both calculations
with the $SF$ and ${\bar G}$-models (dashed) are close to each
other. A slight excess of cross-sections for SF-functions at low
energies results from their extended "tails". Instead, a weak
relative increase in cross-sections at higher energies for the
$\bar G$-functions of the target-nuclei is due to their larger
values in the inner nuclear region, which leads to the earlier
"activation" of absorption as compared to the SF-model.

        The cross-section calculations in the momentum representation
with utilizing the realistic $\rho_{SF}$-densities and the
approximate correction $P_a(R)$-function (\ref{c15}), and the
related approximate profiles (\ref{c16}) and (\ref{c21}) show (see
Fig.4) a slight difference $\approx 2\%$ of such calculations
("bold" points) from the exact ones (solid curves). \\

{ {\large {\bf Table  2.}~~{\it Parameters of the symmetrized
Fermi density distributions $\ro_{SF}(r,c,d)$ of the point-like
nucleons in nuclei
$\cite{BL}$.} \\
} {\large
\begin{center}
\begin{tabular}{|l|c|c|c|c|c|c|}
\hline \hline ~ Nucleus~ & ~$^{12}$C & ~$^{16}$O & ~$^{24}$Mg &
~$^{28}$Si & ~$^{32}$S & $^{40}$Ca
\\
\hline \hline
~$c,~fm$~ & 2.275 & 2.624 & 2.984 & 3.134 & 3.291 & 3.593\\

~$d,~fm$~ & 0.393 & 0.404 & 0.484 & 0.477 & 0.520 & 0.493\\
\hline
\end{tabular}
\end{center}
} }

%%%%%%%%%%%%%%%%%%%%%%%%%%%%%%%%%%%%%%%%%%%%%%%%%%%
\begin{figure}
\begin{center}
\psfig{file=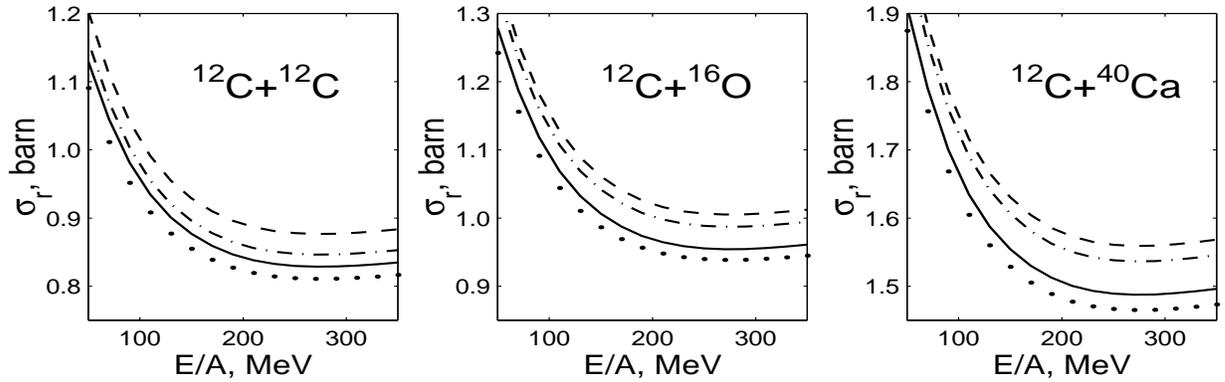,height=5cm,width=.95\linewidth}
\end{center}
\caption{Dependence of cross-sections on the point-like density
parameters obtained from $eA$-scattering data. Solid curves --
with parameters from Table 2, dashed curves -- with parameters
given by eq.(4.4); dashes-dots -- parameters from [37]; bold
circles -- calculations in the momentum representation using
eq.(3.20) (see the text).}
\end{figure}
%%%%%%%%%%%%%%%%%%%%%%%%%%%%%%%%%%%%%%%%%%%%%%%

%\newpage
\section {Calculations and discussion}
\setcounter {equation}0

{\bf {4.1~  On the use of nuclear densities from $eA$-scattering}}\\

\noindent
The convolution integrals (\ref{a7}) and (\ref{a12}) depend on profiles of
both nuclei composed of point-like densities $\ro(s)$, while the transformed
integrals (\ref{a19}), (\ref{a20}) include only one density $\ro(s)$ whereas
the other density is the nuclear one $\rho(s)$. So the problem of obtaining
point-like densities of nuclei is not elucidated completely, and we
consider how they can be extracted from existing data. In general, one can
calculate them in the framework of nuclear models. However, from the outset
we posed the task of using primarily the data of other experiments, for
instance, the data on nuclear charge form factors.  In this case we consider,
so far, that the realistic point-like densities have the form $\ro_{(S)F}(r)$.

The first method \cite{BL} is based on the representation of a nuclear
("experimental") form factor in the form like (\ref{a15}) as follows:
%(4.1)
\begin{equation}\label{d1}
F(q)\,=\,F_P(q)\,F^\circ(q).
\end{equation}
\noindent
Here $F^\circ(q)$ is the form factor of a nucleus with point-like nucleons
and $F_P(q)$ is the proton form factor presented by the dipole formulae
which can be approximated at small momentum by the Gaussian function
\\
%(4.2)
\begin{equation}\label{d2}
F_P(q)\,=\,\Bigl(1+\frac{q^2 r^2_{0\,rms}}{12}\Bigr)^{-2}\simeq
\exp(-q^2 r^2_{0\,rms}/6).
\end{equation}
\noindent
Then, the obtained $F^\circ(q)$ is analyzed within the model-independent
method to obtain a point-like density $\ro(r)$ as a sum of the
$\ro_{SF}(r)$-function with its derivatives multiplied by the fitted
coefficients. The latter reproduce the so-called radial variations
of densities. In this procedure, every $F^\circ(q)$ with the respective
trial density $\ro(r)$ is calculated in the high-energy approximation
\cite{PLP}, \cite{LPP}, the analytical method which gives results in
close agreement with numerical solutions of the Dirac equation. In Table 2
we reproduce part of data from \cite{BL}, namely, the radii $c$ and
diffuseness $d$ parameters of densities $\ro_{SF}(r,c,d)$ excluding the
radial variations which play an important role only at large $q$. In
\cite{BL}, the employed proton $rms$-radius $r^2_{0\,rms}=0.658\,fm^2$
slightly differs from $r^2_{0\,rms}=0.65\,fm^2$, the matter $rms$-radius,
which was used in calculations of the double-folding potentials \cite{SL}.
However, by this reason the point-like densities from \cite{BL} can be
related to the nucleon distributions $\ro_{SF}(r)$ rather than to the
proton ones$\footnote{~If one assumes the relation between proton and
neutron densities to be $\ro_N(r)=(N/Z)\ro_Z(r)$ and takes $rms$-radii
of densities of nuclei consisting of nucleons $R_{rms}^2$, the point-like
nucleons ${\cal R}_{rms}^2$, and protons $\langle r^2\rangle_P=0.76\,fm^2$
and neutrons $\langle r^2\rangle_N=-0.11\,fm^2$ (see \cite{SL}), then it
follows from $R_{rms}^2={\cal R}_{rms}^2 + \langle r^2\rangle_P +
\langle r^2\rangle_N = {\cal R}_{rms}^2 + \langle r^2\rangle$ that the
$rms$-radius of a nucleon is $\langle r^2\rangle=\langle r^2\rangle_P +
\langle r^2\rangle_N = 0.65\,fm^2$.}$.

The other method \cite{FI} of obtaining $c$ and $d$ parameters of
$\ro(r,c,d)$ is based on approximate analytic calculations of
$r^n$-moments of densities $\rho_F(r,R,a)$, where the latter is given
by its folding form like (\ref{a14}). The obtained explicit results for
moments are compared with those obtained by using the standard form
$\rho_F(r,R,a)$ and lead to
\\
%(4.3)
\begin{equation}\label{d4}
c\,=\,R\el[1\,+\,{1\over 3}{\Bigl({r_{0\,rms}\over R}\Bigr)}^2\er], \qquad
d\,=\,a{\el[1\,-{1\over 2}{\Bigl({r_{0\,rms}\over \pi a}\Bigr)}^2\er]}.
\end{equation}
\\
In evaluations, terms of orders higher than $(\pi a/R)^2$ and
$r_{0\,rms}^2/9c^2$ were neglected.

\noindent If one inserts parameters $R$ and $a$ of $\rho_{SF}(r,R,a)$ from
Table 1 and $r_{0\,rms}^2=0.658\,fm^2$ into equations (\ref{d4}), and
compares the obtained values with the respective parameters from Table 2,
one discloses the former to be smaller as compared to those in Table 2
(about 1$\%$ for $c$ and not more than 10$\%$ for $d$). The effect of this
discrepancy on the corresponding total reaction cross sections is shown
in Fig.4. The solid curves are calculations with parameters $c$ and $d$ from
Table 2. The dashed curves are with using $c$ and $d$, estimated by
eq.(\ref{d4}) at $R$ and $a$ from Table 1. The dash-dotted curves show the
cross-sections for $c$ and $d$ from \cite{AFS} ($^{12}C$: 2.1545, 0.425;
$^{16}O$: 2.525, 0.45; $^{40}Ca$: 3.60, 0.523), where they were used
in calculations of the real parts of nucleus-nucleus folding potentials
to explain the elastic scattering cross-sections at energies of about
10 MeV/nucleon. In all the cases we take $r^2_{N\,rms}=0.658\, fm^2$. One
can see that for every set of colliding nuclei differences between the
respective cross-sections occur in the limits of $\approx 6\div 10\%$.
Nevertheless, we incline to believe that a more rigorous method of obtaining
point-like density parameters is to analyze the $F^\circ(q)$ form factors
of nuclei, and thus it is important to make up Tables of the respective
densities.\\

\noindent
{\bf {4.2~ On establishing NN-interaction radius}}\\

It was shown in Sec.2 that at the same radii of the nucleon and
NN-interaction $r_{0\,rms}^2=r_{N\,rms}^2$ the convolution
integral is reduced to a simpler form (\ref{a20}) which contains
only thickness functions of nuclear matter distribution of a
target-nucleus and the point-like density of a projectile nucleus.
As to the NN-interaction parameter $a_N^2=(2/3)r_{N\,rms}^2$, it
is known from the scattering data of free nucleons to give
$r_{N\,rms}^2$ in the limits of $0.63\div 0.69\,fm^2$. At the same
time, in the dipole formula, the nucleon $rms$-radius
$r_{0\,rms}^2$ was used as $0.658\,fm^2$ \cite{BL}, and in
calculations of the double-folding potential \cite{SL} it was
taken to be $0.650\,fm^2$. The effect of their diffferences on the
reaction cross-sections is seen from Fig.5. In calculations,
convolution integrals were taken in the form (\ref{c19}). The
parameters $c$ ¨ $d$ of the point densities for $^{12}C$,
$^{16}O$, $^{40}Ca$ are given in Table 2, and for $^{208}Pb$ they
were evaluated using eq.(\ref{d4}). For the projectile nucleus
$^{12}C$ we used $\bar G$-profile adjusted to the respective
thickness SF-density in the same manner as for Fig.3. It is seen
from Fig.5 that the obtained cross-sections, in fact, coincide to
each others. Therefore, the study of total cross-sections does not
allow us to distinguish between $rms$ radii of a nucleon and the
NN-interaction. Moreover, one should bear in mind that the
amplitude of scattering of free nucleons and its parameter $a_N$
can differ from those scattered in nuclear medium. \\

%%%%%%%%%%%%%%%%%%%%%%%%%%%%%%%%%%%%%%%%%%%%%%%%%%%
\begin{figure}
\begin{center}
\psfig{file=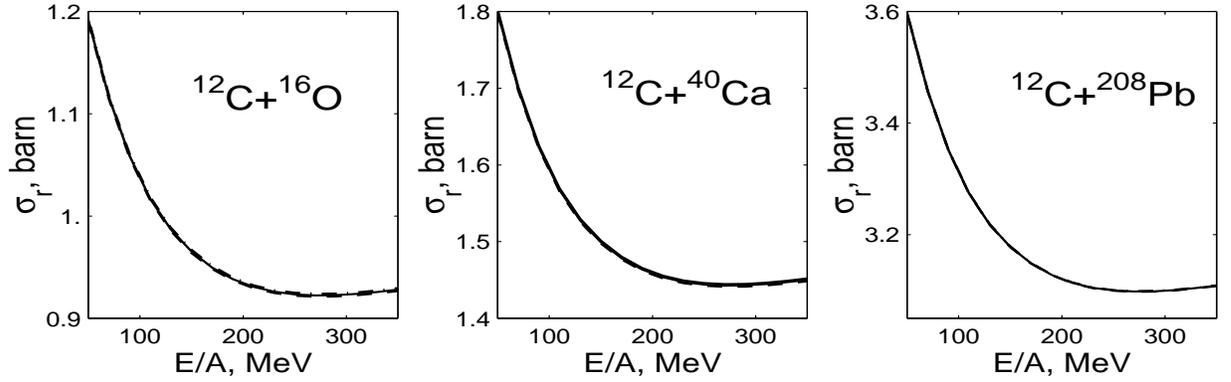,height=5cm,width=.95\linewidth}
\end{center}
\caption{Dependence of cross-sections on the NN-interaction
radius: $r_{N\,rms}^2=0.658\,fm^2$ (solid curves), $0.630\,fm^2$
(dots), $0.69\,fm^2$ (dash-dots). Parameters $c$ and $d$ for
$^{16}O$ and $^{40}Ca$ are in Table 2, and for $^{208}Pb$ they are
done by eq.(4.4) with $R$ and $a$ from Table 1. For $^{12}C$ the
$\bar G$-profile is the same as in Fig.3.}
\end{figure}
%%%%%%%%%%%%%%%%%%%%%%%%%%%%%%%%%%%%%%%%%%%%%%%

\noindent
{\bf {4.3~ Influence of the trajectory distortion}}\\

In the repulsive Coulomb field the trajectory of an incident nucleus
deflects from the scattering center, which results in decrease of the total
reaction cross-section. This effect was taken into account in \cite{VP} by
replacing, in the phase $\chi(b)$, the impact parameter $b$ by the distance
of the turning point $b_c$ in the Coulomb field
%(4.4)
\begin{equation}\label{d6}
 b \,\rightarrow \, b_c={\bar a} + \sqrt{{\bar a}^2+b^2},
\end{equation}
where ${\bar a} = Z_pZ_t e^2/2E_{c.m.}$ is the half-distance of the closest
approach in the field $Z_pZ_t e^2/r$ at $b=0$. Later, the procedure of
exchanging $b$ by $b_c$ was also used for the nuclear part of the
phase $\Phi_N(b)$ in calculations of differential cross-sections of
elastic scattering \cite{VZ}, and, in general, it proved to be correct
(see, for instance, \cite{LZ2}). In addition, at peripheral collisions
one can account for a contribution of the real part V(r) of an attractive
nuclear potential, which brings the Coulomb trajectory closer to the
target-nucleus. If one assumes the region $b\geq R_s = R_p+R_t+(a_p+a_t)\ln2$
to be the main for elastic scattering where the nuclear densities overlap
less than 1/3 of their values in the center \cite{SL}, then the influence of
the "tail" of the nuclear potential can be taken into account by exchanging
%(4.5)
\begin{equation}\label{d7}
b \,\rightarrow \, {\ti b}_c=\ti a + \sqrt{{\ti a}^2+b^2},
\end{equation}
where $\ti a =\bigl (Z_pZ_te^2\,-\,R_s\,|V(R_s)|\bigr )/2E_{c.m.}$.
A more refined way of inclusion of nuclear distortion was elaborated
in \cite{BS} and applied in a series of works (see, for example, \cite{MHC}).
Nevertheless, if the optical potential itself is obtained by numerical
fitting to experimental data, then the use of its real part to correct
the Coulomb trajectory in calculations of the reaction cross-sections
$\sigma_R$ loses its meaning. Indeed, in these cases the data on $\sigma_R$
are usually included into the fit procedure, or, if they are not available,
then they themselves are calculated on the basis of the $S_l$-matrix
elements obtained by fitting the differential cross-sections of elastic
scattering only. Often these "calculated" $\sigma_R$ data are called
"experimental" cross-sections. So the use of the nuclear trajectory
distortion is meaningful only for construction of eikonal phases of distorted
waves when calculating direct inelastic and nucleon removal reactions.
Another situation is when the real part of the nucleus-nucleus
potential is calculated, for example, using the double-folding method.
Then it is reasonable to calculate both the differential and total
cross-sections in the Glauber-Sitenko approach taking into account the
trajectory distortion by both the Coulomb and nuclear field.

%%%%%%%%%%%%%%%%%%%%%%%%%%%%%%%%%%%%%%%%%%%%%%%%%%%
\begin{figure}
\begin{center}
\psfig{file=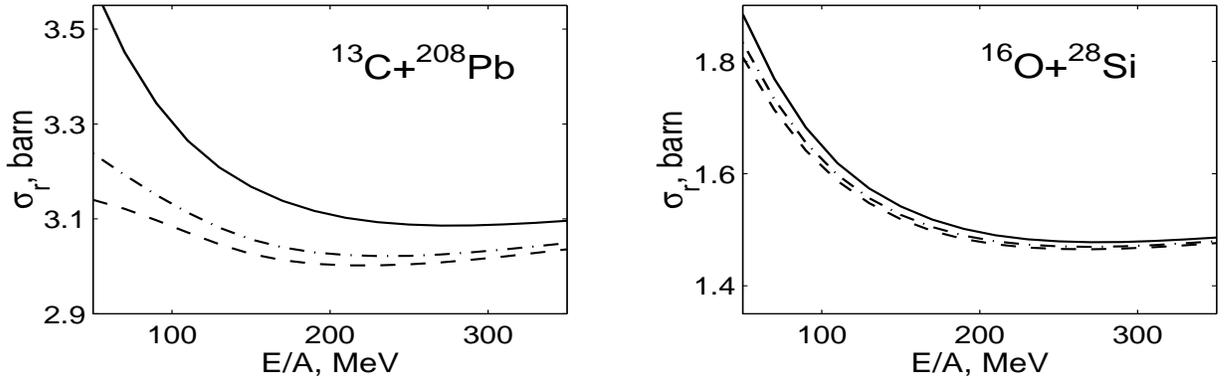,height=5cm,width=.95\linewidth}
\end{center}
\caption{Trajectory distortion effects on cross-sections. Solid
curves -- no distortions; dashes -- only the Coulomb distortion
included; dashes-dots -- effect of the Coulomb and nuclear
distortion. For parameters see the text.}
\end{figure}
%%%%%%%%%%%%%%%%%%%%%%%%%%%%%%%%%%%%%%%%%%%%%%%

  Figure 6 demonstrates the calculations by (\ref{c18}) of the total
cross-sections for $^{13}C+ ^{208}$Pb and $^{16}$O$+^{28}$Si
without including trajectory distortion (solid curves), with the
Coulomb distortion only, using the method (\ref{d6}) (dotted), and
with the Coulomb and nuclear distortion by the method (\ref{d7})
(dash-dotted). In the last case one should specify the parameters
of the nucleus-nucleus potentials, and for $^{13}$C$+^{208}$Pb we
took them from \cite{B} (the "C-potential" at 390 MeV), and for
$^{16}$O$+^{28}$Si from \cite{Satch} (the "E-potential" at 215.2
MeV). In computing phases, we used for $^{13}C$ the parameters of
the $^{12}C$ point-like density from Table 2. The parameters of
$^{16}O$ are also given in Table 2. For $^{208}Pb$ and $^{28}Si$,
the parameters were computed with the help of (\ref{d4}) using the
values $R$ and $a$ from Table 1. As expected, the inclusion of
Coulomb distortion leads to appreciable corrections, of the
order of 10%, of total cross-sections for  heavy nuclei at energies of
100 MeV/nucleon and less, which is beyond experimental errors. As the
collision energy increases, these corrections diminish and for lighter
targets-nuclei (the right-hand part of Fig.6) they reach about 2%, so that
they can be neglected. Contribution of the nuclear distortion is poor in
comparison with the Coulomb one for $^{208}Pb$ and comparable with this
for the reaction on the $^{28}Si$-target, although in the latter case both
the effects give small contributions. Besides, it is necessary to bear in
mind that the real part of the nuclear potential decreases with increasing
energy, but this was not taken into account.\\

\noindent
{\bf {4.4~ In-medium effects}}\\

 In the microscopic approach we deal with the total cross-section
$\sigma_{NN}(\varepsilon_{lab})$  of the NN-scattering of free nucleons.
This cross-section depends on energy, and thus defines the main dependence
of the nucleus-nucleus cross-section on the collision energy $E_{lab}=
\varepsilon_{lab}\,A_p$. We have taken the parametrization of $\sigma_{NN}
(\varepsilon_{lab})$ from \cite{CG} in the energy interval $\varepsilon_
{lab}=10\,MeV \div \, 1\,GeV$. More generally, one should take into account
the in-medium effect on the nucleon-nucleon interactions in nuclear matter.
Usually, for this aim the cross-section is multiplied by the factor $f_m$,
and therefore, in the convolution integral the isospin averaged cross-section
(\ref{d11}) is exchanged as follows:
%(4.6)
\begin{equation}\label{d8}
\sigma_{np}\Rightarrow\sigma_{np}\times f_m(np), \qquad
\sigma_{pp}=\sigma_{nn}\Rightarrow\sigma_{nn}\times f_m(nn).
\end{equation}
\noindent
The factors $f_m(np)$ and $f_m(nn)$ depend of the nucleon energy $\varepsilon_
{lab}=E_{lab}/A$ and on the density of nuclear matter. The problem of
the in-medium corrections of the NN-forces has been investigated in many
works. So in \cite{LM}, based on the Dirac-Bruckner approach for nuclear
matter, numerical calculations were made of the total NN cross-sections.
Parameterization of these calculations was given in \cite{CFS} in analytic
form for the correction factors
\\
%(4.7)
\begin{equation}\label{d12}
f_m(np)=\frac{1+20.88\,\,\varepsilon_{lab}^{\,0.04}\,\,\rho^{\,2.02}}
{1+35.86\,\,\rho^{1.90}}, \qquad
f_m(nn)=\frac{1+7.772\,\,\varepsilon_{lab}^{\,0.06}\,\,\rho^{\,1.48}}
{1+18.01\,\,\rho^{\,1.46}}.
\end{equation}
\\
Here the energy is given in $MeV$, and the density in $fm^{-3}$.
One can see that for free nucleons when $\rho=0$, we have
$f_m(np)=f_m(nn)=1$, and as the density increases these factors
$f_m$ vanish, as well as the corresponding effective
cross-sections$\footnote {~In the microscopic models of the
double-folding potentials the same problems arise when
constructing the energy and in-medium dependence of NN-potentials.
However, the advantage of the Glauber-Sitenko approach is that the
main dependence of energy is already included in parametrization
of the total $\sigma_{NN}$ cross-sections of free nucleons.}$. It
is difficult to compute the convolution integrals with correction
factors in the form (\ref{d12}) including the dependence on $r$,
and we limit ourselves to qualitative estimations only. In Fig.7
are shown such calculations when the values of nuclear densities
in (\ref{d12}) are assumed to be constants $\rho=\bar\rho=
{\bar\rho}_p+{\bar\rho}_t$ for every region where the colliding
nucleons can be. Then, denoting by $\rho_\circ=\rho_p(0)+\rho_t(0)$ the
net density in the centers of colliding nuclei, we show the
calculated cross-sections when the $f_m$ factor are taken for free
nucleons (at $\bar\rho$=0, solid lines), and also for nucleons in
the medium at $\bar\rho$=(1/20)$\rho_\circ$ (bold lines), \,
(1/6)$\rho_\circ)$ (dots), \, (1/3)$\rho_\circ$ (dash-dotted lines),
\, $\rho_\circ$ (dashed curves). It is seen that the inclusion of
the in-medium factors $f_m$ can diminish the total reaction
cross-section by 4-7\% and that the dependence on the density
turns out to be strongly nonlinear.

%%%%%%%%%%%%%%%%%%%%%%%%%%%%%%%%%%%%%%%%%%%%%%%%%%%
\begin{figure}
\begin{center}
\psfig{file=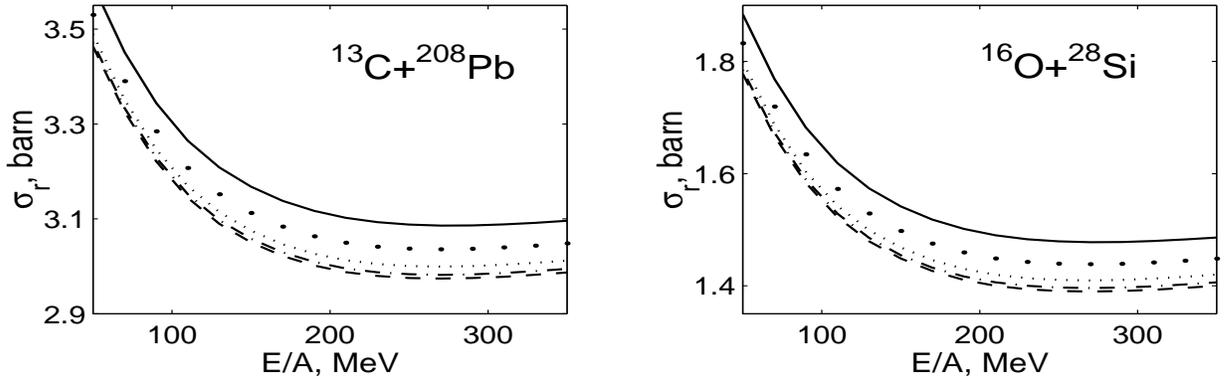,height=5cm,width=.95\linewidth}
\end{center}
\caption{In-medium effect on cross-sections. Solid curves -- no
in-medium dependence ($\bar\rho$=0); bold dots are when the density
in the overlap region is taken as 1/20$\rho_\circ$ with
$\rho_\circ=\rho_p(0)+\rho_t(0)$ expressed through densities at the
centers of nuclei; dots -- for overlap with 1/6$\rho_\circ$;
dashes-dots -- for 1/3$\rho_\circ$; dashes -- at $\rho_\circ$.}
\end{figure}
%%%%%%%%%%%%%%%%%%%%%%%%%%%%%%%%%%%%%%%%%%%%%%%

\section {Conclusions, comparison with experimental data}
\setcounter {equation}0

\hspace*{0.6cm}1.~It is shown that the initial expression for the microscopic
eikonal phase can be represented in a convenient form where one of two
density thickness functions of point-like nucleons in nuclei transforms into
the profile of a nuclear density obtained from independent experimental data,
e.g., from nuclear form factors. Ultimately, it turns out that in calculations
of nucleus-nucleus cross-sections it is not necessary to introduce free
parameters but, instead, to base on the known experimental data, only.

2. The main problem of the microscopic approach is to calculate eikonal
scattering phases. For this aim, many people use the simplest nuclear
densities in the form of Gaussian functions. Instead, here we demonstrate
how one can use the realistic density distributions in the form of
(symmetrized) Fermi functions, whose parameters are known for many nuclei.
In this case, all functions in the eikonal integrals turn out to be given
in the explicit form, which highly simplifies numerical calculations.

3.~Two methods of obtaining the radius $c$ and diffuseness $d$
parameters of the point-like density $\ro_{SF}(r,c,d)$ are
presented. One of them deals with nuclear form factors, and the
other obtains these parameters from the known nuclear densities
$\rho_{SF}(r,R,a)$. It turns out that the difference between the
cross-sections calculated by using these two sets of parameters
can achieve the value beyond the bars of typical experimental
cross-sections. We conclude that a more justified method is that
based on the analysis of form factors of nuclei obtained by
subtracting the nucleon form factor and, if necessary, the form
factor of the nuclear center-of-mass motion, as was carried out,
for example, in [29].

4.~In calculation, it was established that the $rms$ radii of the
nucleon-nucleon interaction $r_{N\,rms}^2$ and of the nucleon itself
$r_{0\,rms}^2$ may be considered to be equal. The slight difference between
them is in the range of accuracy of their evaluation, and this does not
affect practically the results of calculations of the nucleus-nucleus total
reaction cross-sections. At the same time at $r_{N\,rms}^2 =r_{0\,rms}^2$
the convolution integrals take highly simple forms (\ref{a19}), (\ref{a20})
where only two profile functions overlap, one is for the nuclear density
and the other is for the point-like density. Next, the convolution integral
takes a simple one-dimensional form.

5.~In many typical cases one should take into account a Coulomb trajectory
distortion by means of the formal replacing, in the phase $\chi(b)$, the
impact parameter $b$ by $b_c$, according to (\ref{d6}). Additional
distortion of the trajectory by the tail of a nuclear potential is
not reasonable because its parameters are to be fitted, in particular,
to the same data on total cross-sections.

6.~The question whether the in-medium factor $f_m$ for corrections of
the NN-cross-sections should be taken into consideration remains open for
us till now. The estimations show that at intermediate energies this
factor does not change substantially the nucleus-nucleus total cross-sections.
Nevertheless, the use of the above given factors $f_m(np)$ and $f_m(nn)$
from \cite{CFS} hinder the usual calculations. Other authors faced similar
obstacles, for example, when solving a simpler problem of the $pA$ scattering
data analysis \cite{VRCC}. Besides, in both these works, the Gaussian functions
were chosen as the basic nuclear densities, while their replacing by the
realistic ones can change the conclusions about the structure of the factors
$f_m$ as compared to those given by eq.(\ref{d12}). For example, successful
agreement of calculations with the data on total reaction cross-sections
for a series of reactions pA, $\alpha A$ and ${^{12}C}+{^{12}C}$ was obtained
in \cite{TWC} with realistic densities and with the factors $f_m$ differing
in form from those given above $\footnote {~In calculations of the real part of the double-folding
nucleus-nucleus potential, the in-medium effect on the NN-potential is
parametrized by simpler factors $f_m$ in the form of the step and
exponential functions of densities $\rho(r)$ (see., e.g., \cite{KSO}).}$.

7.~Our calculations show that in the case of relatively light incident
nuclei, it is more profitable to use the convolution integral (\ref{c18})
with the thickness function $\ro_p(s)$ for the point-like density of the
projectile nucleus in the form of the modified Gaussian $\bar G$-function.
The latter is determined with the help of (\ref{c6}) and (\ref{c7}) by means
of parameters of realistic SF-function taken, for example, from Table 2.
In the case of heavier incident nuclei it is advantageous to use the
convolution integral (\ref{a19}) where one can insert the known explicit
form of the Fourier-Bessel profiles of realistic SF-densities (\ref{c23}) for
both the nuclei. In all calculations the Coulomb distortion of trajectory
should be first of all taken into account.

8.~We emphasize that in the existing literature no attention is focused
on the problem of correct use of the initial formula for the convolution
integral. From the above discussion it is evident that if the finite radius
of the NN-interaction ($a^2_N\neq 0$) is taken into account explicitly, both
the densities must be taken as densities of nuclei for the point-like
nucleons. Further, when the rms-radii of the nucleon and the NN-interaction
are equal to each other, the NN-interaction factor disappears in the
convolution integral, but one of the point-like densities is transformed
to the nuclear one. However, in this case, the absence of the NN-factor
$f(\xi)$ does not mean that one uses the zero-range approximation. The
confusion arises also, when one assumes $a^2_N=(2/3)r^2_{N\,rms}=0$ in
(\ref{a7}) and calls this case the zero-range approximation, while at
the same time takes for both the densities the table data, i.e., just
the nuclear densities, instead of the point-like ones.

%%%%%%%%%%%%%%%%%%%%%%%%%%%%%%%%%%%%%%%%%%%%%%%%%%%
\begin{figure}
\begin{center}
\psfig{file=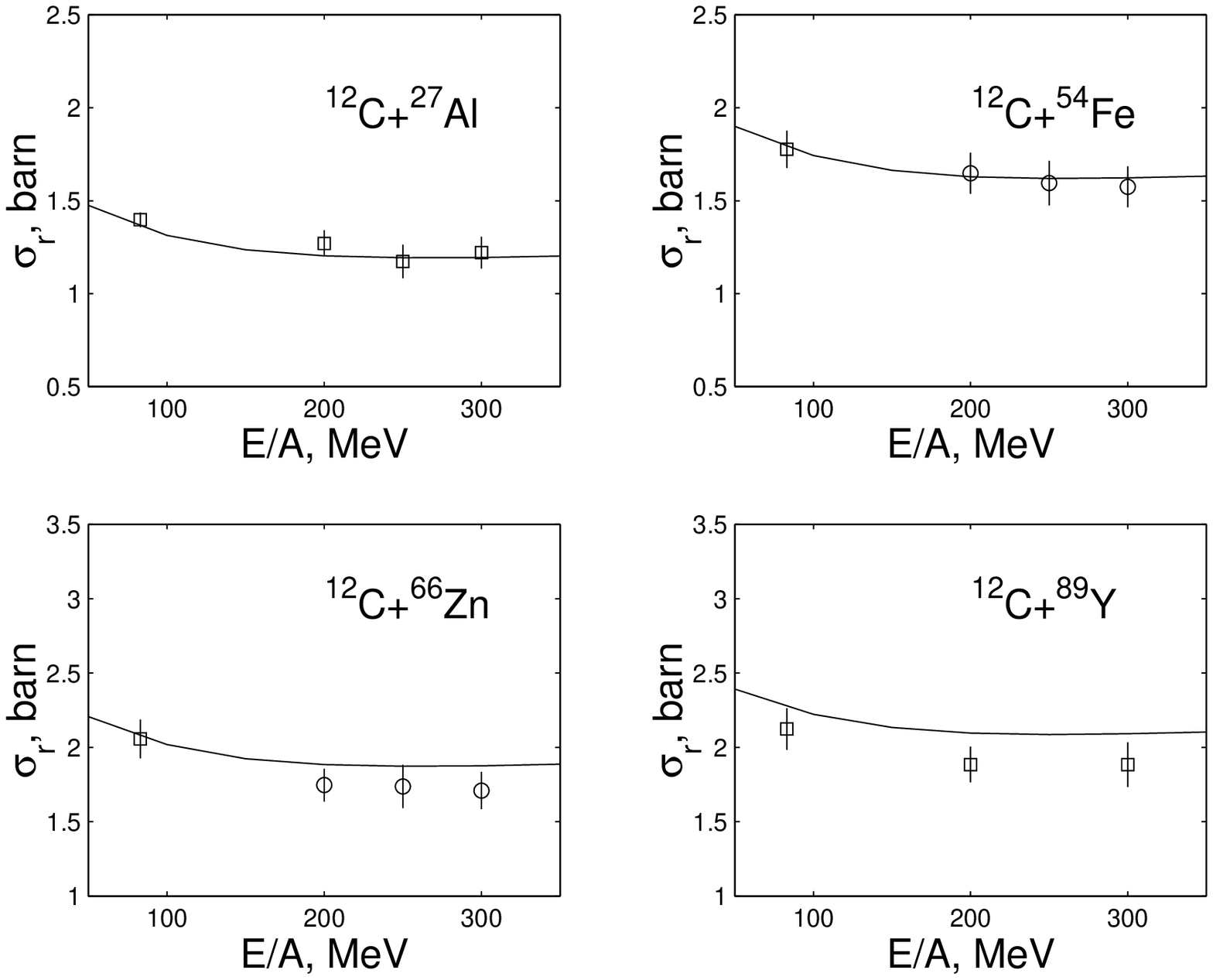,height=13cm,width=.95\linewidth}
\end{center}
\begin{center}
\psfig{file=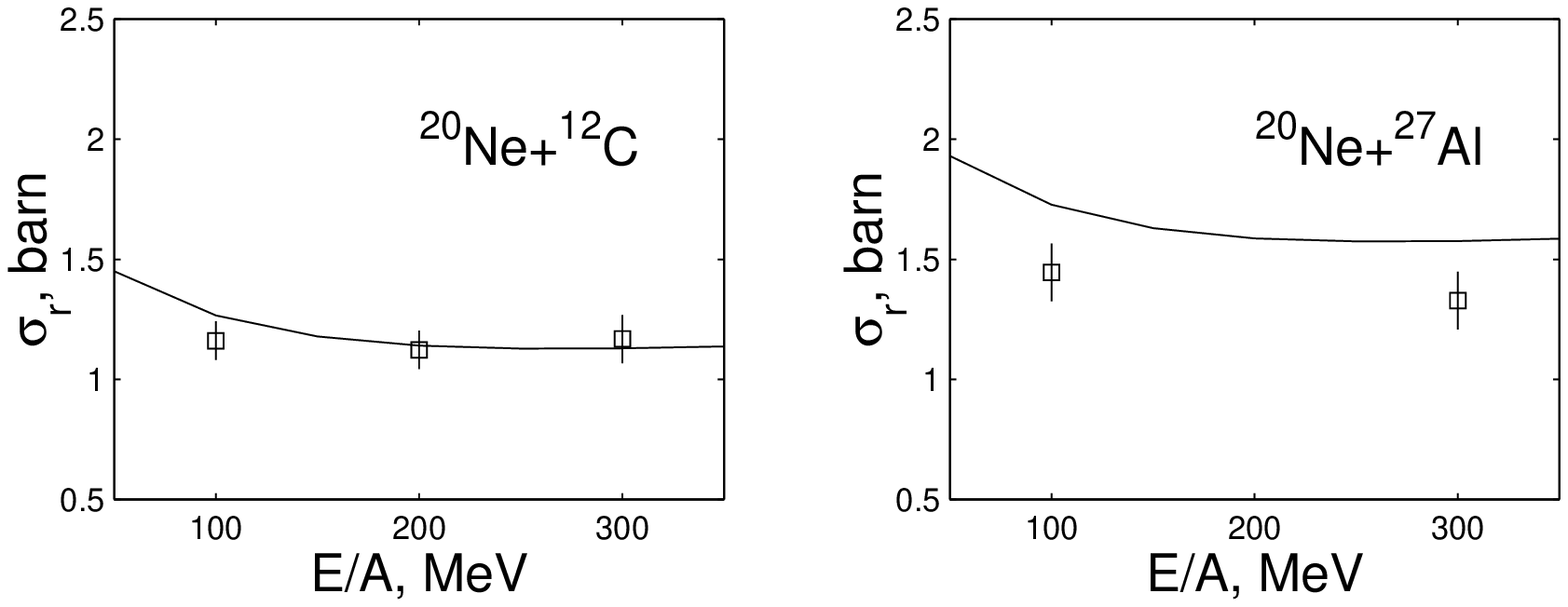,height=6cm,width=.95\linewidth}
\end{center}
\caption{Calculated cross-sections compared with experimental data
from [48]. Parameters are from the $eA$-scattering data (see in
the text). Only the Coulomb distortion is included.}
\end{figure}
%%%%%%%%%%%%%%%%%%%%%%%%%%%%%%%%%%%%%%%%%%%%%%%

In Fig.8a,b, our calculations are compared with the experimental data
collected in [49]. The SF-density parameters are given in Table 1. For the
density $\rho_{SF}^\circ$ the parameters $c$ and $d$ of the projectile
$^{12}C$ are given in Table 2, and for the incident nucleus ${^{20}Ne}$
they are calculated with the help of (\ref{d4}) and the data from Table 1.
The convolution integral has been taken in the form of (\ref{c18}). Only
the Coulomb trajectory distortion was taken into account. Thus, no free parameters have been
introduced. One can see that in all cases a good agreement with experimental
data is observed, except for ${^{12}C}+{^{89}Y}$ and ${^{20}Ne}+{^{27}Al}$.
In the latter cases, some discrepancies can be due to the fact that the
parameters for densities of the odd nuclei ${^{89}Y}$ and ${^{27}Al}$ were
obtained from $eA$ form factors (for references see [28]) with the help of
formulae for spinless nuclei. However, if one introduces the in-medium factor
$f_m$, the mentioned disagreements can be removed. Nevertheless, we
believe that first of all one should improve the data on geometric parameters
of the given nuclei. The other remark is that calculations of cross-sections
in \cite{Kox} with the help of Gaussian functions \cite{Ka}, which reproduce
the behavior of "tails" of densities, give enhanced values which are beyond
the experimental bars shown in Figs.8a,b. This is conceivable that
in \cite{Kox} they took the nuclear densities instead of the point-like ones,
as it was mentioned above. On the other hand, calculations in \cite{AT} with
the uniform density distributions for the target-nuclei and Gaussian functions
for the projectiles give underestimations of the cross-sections for
${^{20}Ne}+{^{12}C}$ and ${^{12}C}+{^{27}Al}$, in spite of that they,
probably, use not the point-like but nuclear densities for estimations
of $rms$-radii needed to get the step radius $R_u$ in (\ref{c3}). So, this
result confirms that shown in Fig.3.

In conclusion, the authors would like to thank the Infeld-Bogolubov
Foundation for its support, and E.B.Z. acknowledges the support of the
Russian Foundation for Basic Research (grant 0001-006-17).

%\vspace*{1cm}


\begin{thebibliography}{99}
%\vskip -5 mm
\itemsep -2mm
%[1]
\bibitem {G} ~R.J.Glauber, {\it Lectures on Theor. Phys.}, {\bf 1}
              (Interscience, New York, 1959).
%[2]
\bibitem {S} ~€.G.Sitenko, Ukr.Fiz.J. {\bf 4} 152 (1959);
%[3]
\bibitem {Sat} ~G.D.Satchler, {\it Direct nuclear reactions} (Clarendon,
                Oxford, 1983).
%[4]
\bibitem {LSZ} ~V.K.Lukyanov, B.~Slowinski, E.~Zemlyanaya, Yad.Fiz. {\bf 64}
               1349 (2001);\\
               ~V.K.Lukyanov, B.~Slowinski, E.~Zemlyanaya, Phys.At.Nucl.{\bf 64}
%[5]           1273 (2001)
\bibitem {W} ~J.-S.Wan a.o., Kerntechnik {\bf 63} 167 (1998).
%[6]
\bibitem {TW} ~L.W.Townsend, J.W.Wilson, Rad.Res. {\bf 106} 283 (1986).
%[7]
\bibitem {FST} ~S.Fernbach, R.Serber, T.B.Teylor, Phys.Rev. {\bf 75} 1352
              (1949).
%[8]
\bibitem {Czyz} ~W.Czyz, L.S.Maximon, Ann.of Phys. {\bf 52} 59 (1969).
%[9]
\bibitem {Formanek} ~J.Formanek, Nucl.Phys.B {\bf 12} 441 (1969).
%[10]
\bibitem {T} ~I.Tanihata, J.Phys.G {\bf 22} 157 (1996).
%[11]
\bibitem {BCH} ~C.A.Bertulani, L.F.Cano, M.S.Hussein, Phys.Rep. {\bf 226}
                281 (1993).
%[12]
\bibitem {KKF} ~O.M.Knyaz'kov, I.N.Kukhtina, S.A.Fayans, Fiz.Elem.Chast.{\&}
                At.Yad.{\bf 28} 1061 (1997);\\
                ~O.M.Knyaz'kov, I.N.Kukhtina, S.A.Fayans, Sov.J.Part.Nucl.
                {\bf 28} 418  (1997).
%[13]
\bibitem {VZ} ~A.Vitturi, and F.Zardi, Phys.Rev.C {\bf 36} 1404 (1987).
%[14]
\bibitem {MHC} ~Moon Hoe Cha, Phys.Rev.C {\bf 46} 1026 (1992).
%[15]
\bibitem {AZW} ~C.E.Aguiar, F.Zardi, A.Vitturi, Phys.Rev.C {\bf 56} 1511
               (1997).
%[16]
\bibitem {TJAK} ~J.A.Tostevin, R.C.Johnson, J.S.Al-Khalili, Nucl.Phys.A
               {\bf 630} 340c (1998).
%[17]
\bibitem {BBS} ~G.F.Bertsch, B.A.Brown, H.Sagava, Phys.Rev.C {\bf 39} 1154
               (1989).
%[18]
\bibitem {AAV} ~G.D.Alkhazov, V.V.Anisovich, P.E.Volkovitskiy, {\it Diffractive
interactions of hadrons with nuclei at high energies} (Nauka, Leningrad, 1991)
(in russian).
%[19]
\bibitem {SL} ~G.R.Satchler and W.G.Love, Phys.Rep. {\bf 55} 183 (1979).
%[20]
\bibitem {DK} ~O.D.Dalkarov, V.A.Karmanov, Nucl.Phys.A {bf 445} 579 (1985).
%[21]
\bibitem {Rihan} ~H.M.Fayyad, T.H.Rihan, A.M.Awin, Phys.Rev.C {\bf 53} 2334
                  (1996).
%[22]
\bibitem {Ka} ~P.J.Karol, Phys.Rev.C {\bf 11} 1203 (1975).
%[23]
\bibitem {CG} ~S.Charagi and G.Gupta, Phys.Rev.C {\bf 41} 1610 (1990).
%[24]
\bibitem {ELP} ~Yu.N.Eldyshev, V.K.Lukyanov, Yu.S.Pol', Yad.Fiz. {\bf 16} 506
               (1972);\\
               ~Yu.N.Eldyshev, V.K.Lukyanov, Yu.S.Pol', Sov.J.Nucl.Phys.
               {\bf 16} 282 (1973).
%[25]
\bibitem {SM} ~D.W.L.Sprung, J.Martorell, J.Phys.A {\bf 30} 6525 (1997);
               {\it ibid.} {\bf 31} 8973 (1998).
%[26]
\bibitem {GKLS} ~M.Grypeos, C.Koutroulos, V.Lukyanov, A.Shebeko,
          J.Phys.G {\bf 24} (1998) 1998.
%[27]
\bibitem {BKLP} ~V.V.Burov, D.N.Kadrev, V.K.Lukyanov, Yu.S.Pol',Yad.Fiz.
                  {\bf 61} 595 (1998);\\
                ~ V.V.Burov, D.N.Kadrev, V.K.Lukyanov, Yu.S.Pol',
                Phys.At.Nucl.{\bf 61} 525 (1998).
%[28]
\bibitem {VJV} ~De Vries H., de Jager C.W., de Vries C., At. Data $\&$
                Nucl. Data Tables {\bf 36} (1987) 495.
%[29]
\bibitem {BL} ~V.V.Burov, V.K.Lukyanov, Preprint P4-11098, JINR (Dubna, 1977).
%[30]
\bibitem {LZ} ~V.K.Lukyanov, E.V. Zemlyanaya, J.Phys.G. {\bf 26} 357 (2000).
%[31]
\bibitem {RG} ~I.Gradshtein, I.Ryzhik, {\it Tables of Series, Products and
               Integrals} (H.Deutsh, 1987)(Engl.transl.).
%[32]
\bibitem {AS} ~M.Abramowitz and I.A.Stegun, {\it Handbook of Mathematical
              Functions}, (Nat. Bureau of Standards, Appl. Math. Series,
              1964).
%[33]
\bibitem {GKLS2} ~M.E.Grypeos, C.G.Koutroulos, V.K.Lukyanov, A.V.Shebeko,
                 Fiz.Elem.Chast.{\&} At.Yad.{\bf 32} 1494 (2001);\\
                 ~M.E.Grypeos, C.G.Koutroulos, V.K.Lukyanov, A.V.Shebeko,
                 Phys.Particles{\&}Nucl., {\bf 32} 779 (2001).
%[34]
\bibitem {PLP} ~I.Zh. Petkov, V.K.Lukyanov, Yu.S.Pol', Yad.Fiz.,{\bf 4} 57
               (1966).
%[35]
\bibitem {LPP} ~I.Zh. Petkov, V.K.Lukyanov, Yu.S.Pol', Yad.Fiz.,{\bf 9} 349 (1969).
%[36]
\bibitem {FI} ~S.F$\ddot a$ldt, A.Ingemarssen, J.Phys.G. {\bf 9} 261 (1983).
%[37]
\bibitem {AFS} ~M.El-Azab Farid, G.R.Sachler, Nucl.Phys.A {\bf 438} 525 (1985).
%[38]
\bibitem {VP} ~R.M.DeVries, and J.C.Peng, Phys.Rev.' {\bf 22} 1055 (1980).
%[39]
\bibitem {LZ2} ~V.K.Lukyanov, E.V. Zemlyanaya, Int.J.Mod.Phys.E. {\bf 10}
                no.3, 169 (2001).
%[40]
\bibitem {BS} ~D.M. Brink, and G.R.Satchler, J.Phys.G. {\bf 7} 43 (1981).
%[41]
\bibitem {B} ~M.Buenerd et al., Nucl.Phys.A {bf 424} 313 (1984).
%[42]
\bibitem {Satch} ~G.R.Sachler, Nucl.Phys.A {bf 279} 493 (1977).
%[43]
\bibitem {LM} ~G.Q.Li and R Machleidt, Phys.Rev.' {\bf 48} 1702 (1993);
               {\it ibid.} {\bf 49} 566 (1994).
%[44]
\bibitem {CFS} ~Cai Xiangzhou, Feng Jun, Shen Wenqing, Ma Yugang, Wang Jiansong,
              and Ye Wei, Phys.Rev.' {\bf 58} 572 (1998).
%[45]
\bibitem {VRCC} ~A.de Vismes, P.Roussel-Chomaz, and R.Carstoiu, Phys.Rev.'
                {\bf 62} 064612-1 (2000).
%[46]
\bibitem {TWC} ~R.K.Tripathi, J.W.Wilson, F.A.Cucinotta, Nucl.Instr.$\&$
               Meth.Phys.Res.B {\bf 145} 277 (1998).
%[47]
\bibitem {KSO} ~Dao T.Khoa, G.R.Satchler, W.von Oertzen, Phys.Rev.C {\bf 56}
                 954 (1997).
%[48]
\bibitem {Kox} ~S.Kox et al, Phys.Rev.C {\bf 35} 1678 (1987).
%[49]
\bibitem {AT} ~A.Y.Abul-Magd, M.Talib Ali-Alhinai, Nuovo Cim.A {\bf 110}
                1281 (1997).

\end{thebibliography}
\end{document}